\documentclass[nato,noid,numreferences]{crckbked}

\usepackage{graphicx}
\usepackage{epsf}
\usepackage{amssymb}
\newcommand{\beq}{\begin{equation}}
\newcommand{\eeq}{\end{equation}}
\newcommand{\bea}{\begin{eqnarray}}
\newcommand{\eea}{\end{eqnarray}}
\def\={\;=\;}
\def\+{\;+\;}
\newcommand{\ave}[1]{\langle {#1} \rangle}
\newcommand{\eq}[1]{Eq.~(\ref{#1})}
\newcommand{\eqs}[1]{Eqs.~(\ref{#1})}
\newcommand{\ct}{s_{22}}
\newcommand{\cf}{s_{55}}
\newcommand{\cs}{s_{77}}
\newcommand{\pb}{\bar\psi}

\begin{document}
% The \begin{document} command comes before the \begin{opening}
% command.

\begin{opening}
\title{COLOR SUPERCONDUCTIVITY IN TWO- AND THREE-FLAVOR SYSTEMS AT 
       MODERATE DENSITIES}

%\subtitle{Basic Instructions}
% Uncomment if you want to give a subtitle.

% You can split the title and subtitle by putting 
% two backslashes at the appropriate place. 

\author{Michael Buballa}
\institute{Institut f\"ur Kernphysik, TU Darmstadt\\ 
           Schlossgartenstr. 9, D-64289 Darmstadt, Germany;\\
           GSI Darmstadt\\
           Postfach 110552; D-64220 Darmstadt, Germany}
% If there are more authors at one institute, you should first
% use \author{...} for each author followed by \institute{...}.
% Put 
% \author{and} 
% \institute{}
% before the last author.

\author{and}
\institute{}

\author{Micaela Oertel}
\institute{IPN Lyon\\  
           4, rue Enrico Fermi; F-69622 Villeurbanne C\'edex, France}

\begin{abstract}
Basic features of color superconductivity are reviewed, 
focusing on the regime of ``moderate densities'', which is not 
accessible by perturbation theory. We discuss the standard picture of
two- and three flavor color superconductors and study the 
color-flavor unlocking phase transition within an NJL-type model.
\end{abstract}

\end{opening}

\section{Introduction}

The structure of the QCD phase diagram is one of the most exciting topics
in the field of strong interactions (For reviews see, e.g. 
\cite{Halasz99,Raja99,RaWi00,Alford01,San02}).
For a long time the discussion was 
restricted to two phases: the hadronic phase and the quark-gluon plasma (QGP).
The former contains ``our'' world, where quarks and gluons are confined
to color-neutral hadrons and chiral symmetry is spontaneously broken due to
the presence of a non-vanishing quark condensate $\phi = \ave{\pb\psi}$.
In the QGP quarks and gluons are deconfined and chiral symmetry is
(almost) restored, $\phi\simeq 0$.
 
Although color-superconducting phases were discussed already in the '70s 
\cite{CoPe75,Ba77,Fr78} and '80s \cite{BaLo84}, until quite recently not much 
attention was payed to this possibility.
This changed dramatically after it was discovered that due to 
non-perturbative effects, the gaps which are related to these phases could 
be of the order of $\Delta\sim 100$~MeV \cite{ARW98,RSSV98}, much larger 
than expected from the early perturbative estimates. 
Since in standard weak-coupling BCS theory the critical temperature is given 
by $T_c \simeq 0.57\,\Delta(T=0)$ \cite{FW71}, this also implies
a sizable extension of the color-superconducting phases into the 
temperature direction \cite{PiRi99}.
It was concluded that color-superconducting phases could be relevant
for neutron stars \cite{We99,BGS01} and -- in very optimistic cases -- 
even for heavy-ion collisions \cite{PiRi00}. 

Rather soon after the beginning of this new era, it was noticed that
there is probably more than one color superconducting phase in the 
QCD phase diagram. At large chemical potential, where up, down, and
strange quarks can condense, matter is expected to be in the so-called
color-flavor locked phase \cite{ARW99}, whereas at intermediate densities, 
just above the deconfinement phase transition, we might have a 
two-flavor color superconductor (2SC). Other phases have been
suggested more recently, like crystalline phases in a small window
between the 2SC phase and the CFL phase \cite{RSZ01,ABoR01}
or a CFL phase accompanied by a kaon condensate (CFL + K) 
\cite{Sch00a,BeSc02}. 
In addition, those color or flavor degrees of freedom which do not
participate in the ``standard'' condensates could pair in different,
usually more fragile, channels, thus forming additional phases
\cite{Sch00b,BHO02b,SanS02}. 

In this talk we review some of the basic features of color superconductivity,
mainly focusing on the ``standard'' phases for two and three flavors
and the transition from the 2SC to the CFL phase. 
Very recently (after the Stara Lesna workshop) it has been argued that
in neutron stars there might be no 2SC phase at all, because of the
rather different Fermi surfaces of $u$- and $d$- quarks in charge neutral
matter \cite{AlRa02}. We will briefly comment on this possibility in the 
end of this article.

\section{Diquark condensates}

According to Cooper's theorem any arbitrarily weak
interaction leads to an instability at the Fermi surface which is cured
by the formation of Cooper pairs. At very large densities, where
asymptotic freedom allows to perform the analysis in terms of a
single gluon exchange it can easily be shown that there are indeed
attractive channels and hence QCD matter must be a color superconductor
at these densities. 
However, because of the large number of possible channels related to
the quantum numbers of spin, flavor and color, we can almost be sure
that also in the nonperturbative regime just above the deconfinement phase 
transition some of them will be attractive. 

In general, a diquark condensate may be written as 
$\ave{\psi^T {\cal O} \psi},$
where $\psi$ is a quark field and ${\cal O}$ an operator, acting in color, 
flavor and Dirac space.
It can also contain derivatives, but we will not consider
this possibility here. 
A priori, the only restriction to ${\cal O}$ is provided by the Pauli
principle, which requires that ${\cal O}$ must be totally antisymmetric.
This still leaves many possibilities, and thus the interaction must decide
about the actual condensation pattern. 

As already mentioned, at very large chemical potentials, 
$\mu \gg \Lambda_{QCD}$, $\alpha_s(\mu)$ is small and the problem 
can be (and has been) attacked from first principles
\cite{PiRi99,PiRi00,Son99,SW99,HMSW00}. 
To estimate the range of validity of these calculations we assume
(quite optimistically) that the perturbative regime begins at
\mbox{$\mu \approx 1.5$~GeV}. For two massless flavors this
corresponds to a baryon density \mbox{$\rho_B = 2/(3\pi^2) \mu^3 \approx
30$~fm$^{-3}$}, which is about 175 times nuclear saturation density.
It turns out that the situation is even worse:
In a numerical study Rajagopal and Shuster \cite{RaSh00}
found that (gauge dependent)
higher-order terms can only be neglected if $\mu \gg 10^5$~GeV!

Hence, asymptotic studies, although interesting by themselves, cannot
be trusted down to densities which are present, e.g., in the interior
of neutron stars. 
In this regime one has to rely on effective interactions, like
instanton interactions \cite{RSSV00}, or (local or nonlocal) 4-point 
interactions (``NJL-type models'')
whose structure is also abstracted from the instanton vertex
\cite{ARW98} or purely phenomenological.
This is quite analogous to the Landau-Migdal interaction used to
describe nuclear matter. However, we should be aware of the fact
that there are presently no data to constrain the parameters in
the deconfined phase itself. They are therefore usually fixed in vacuum,
which is clearly a source of big uncertainties.  

Nevertheless, there are good reasons to believe, that a dominant role 
might be played by the Lorentz-invariant scalar ($J=0^+$) condensate,
\beq
    s_{AA'} \= \ave{\psi^T C\gamma_5\,\tau_A\lambda_{A'} \psi}~,
\label{saa}
\eeq
which corresponds to the most attractive channel, both for interactions
with the quantum numbers of a single gluon exchange as well as for 
instanton induced interactions. Here $C$ is the matrix of charge
conjugation, and $\tau_A$ and $\lambda_{A'}$ are the antisymmetric generators
of flavor-$SU(N_f)$ and color-$SU(N_c)$, respectively. 
Throughout this article, we will restrict ourselves to
the physical number of colors, $N_c=3$. Then the $\lambda_{A'}$ denote the
three antisymmetric Gell-Mann matrices, $\lambda_2$, $\lambda_5$ and 
$\lambda_7$, i.e., $s_{AA'}$ is a color anti-triplet.
Concerning the number of flavors we begin with $N_f=2$ in the next
section and we will dicuss $N_f=3$ later on.

\section{Two flavors}

For two flavors ($N_f= 2$), the flavor index in \eq{saa} is restricted to 
$A=2$, describing the pairing of an up quark with a down quark.
In the limit of massless up and down quarks $s_{2A'}$
is invariant under chiral $SU(2)_L \times SU(2)_R$ transformations. 
The three condensates $s_{22}$, $s_{25}$, and $s_{27}$, form a vector in 
color space, which always can be rotated into the $A' = 2$-direction. 
Hence the two-flavor superconducting state (2SC) state can be 
characterized by
\beq
    \ct \neq 0  \quad {\rm and}\quad
    s_{AA'} = 0 \quad {\rm if}\quad (A,A') \neq (2,2) \,.
\label{2sc}
\end{equation}
Since only the first two colors (``red'' and ``green'') participate in
the $\ct$, while the third one (``blue'') does not,
color $SU(3)$ is spontaneously broken down to $SU(2)$. 
As a result five of the eight gluons acquire a mass \cite{Ri00}. 

Like in ordinary BCS theory the pairing produces a gap in the spectrum
of the corresponding quasiparticles, which in the case of massless
quarks is characterized by the dispersion laws
\beq
    E_\mp(\vec p) \= \sqrt{(p\mp\mu)^2 + |\Delta|^2}~,
\eeq    
where $p = |\vec p|$. The gap $\Delta$ is proportional to $\ct$ and
is determined by a gap equation. For local 4-point intercations
the latter takes the form
\beq
    \Delta \={\it const.} \Delta\,\int\frac{d^3 p}{(2\pi)^3}\;
     ( \frac{1}{E_-}\tanh{\frac{E_-}{2T}} +
       \frac{1}{E_+}\tanh{\frac{E_+}{2T}})~,
\eeq 
where {\it const.} contains a coupling constant in the scalar diquark
channel and degeneracy factors.  The result is typically of the order
of 100~MeV \cite{ARW98,RSSV98}.

Until this point we have neglected all other possible condensates, which
might compete or coexist with $\ct$.
Because of the empirical fact that the (approximate) chiral SU(2)
symmetry of the QCD Lagrangian is not respected by the QCD vacuum, it is
natural to ask whether the quark (-antiquark) condensate
\begin{equation}
     \phi \=\ave{\pb \,\psi} \;,
\label{phi}
\end{equation}
persists also in the ground state of QCD matter at finite baryon
density. This question has been addressed first by Berges and
Rajagopal \cite{BeRa99} within a phenomenological NJL-type model.
For massless quarks at $T=0$ they found a first-order phase transition 
from the vacuum state with $\phi\neq 0$ and $\ct = 0$ to a 
high-density phase with $\ct \neq 0$ and $\phi = 0$.  This is
different if there is a small quark mass $m$ which explicitly
breaks chiral symmetry. In this case $\phi$ cannot exactly vanish
above the phase transition and coexists with the diquark
condensate. In fact, just above the phase transition the gaps
related to the two condensates can be of similar magnitude
\cite{BeRa99}.

However, this is not yet the whole story:
At finite density the existence of Lorentz non-invariant expectation values
becomes possible. The most obvious example is of course the density itself,
\beq
     \rho \= \ave{\pb \,\gamma^0\,\psi} \;,
\label{rho}
\end{equation}
which transforms like the time component of a 4-vector.
Together with $\ct$ and $\phi$ this means that color-$SU(3)$, chiral symmetry
and Lorentz invariance are broken in the system. 
Therefore a fully selfconsistent description requires to take 
into account further condensates which are no longer prohibited by 
one or more of the above symmetries. 

It turns out that in general three more condensates should be considered:
First, after color-$SU(3)$ is broken, there is no need for the scalar
and vector densities to be the same for ``red'' and ``blue'' quarks.
Hence, in addition to \eqs{phi} and (\ref{rho}) there could be
condensates of the form   
\beq
     \phi_8 \= \ave{\pb \,\lambda_8\,\psi}
            \= \frac{2}{\sqrt{3}}\,(\phi_r - \phi_b) \;
\label{phi8}
\end{equation}
and
\beq
     \rho_8 \=\ave{\pb \,\gamma^0\,\lambda_8\,\psi}
            \= \frac{2}{\sqrt{3}}\,(\rho_r - \rho_b) \;.
\label{rho8}
\end{equation}
Note that all green quantities are equal to the red ones. Since the scalar densities are closely related to the constituent
quark masses, a non-vanishing $\phi_8$ would mean that the constituent
masses of red and blue quarks, $M_r$ and $M_b$, can differ from each other.

Finally, there could be another diquark condensate of the
form \cite{BaLo84,LaRh99,ABR99}
\beq
     \ct' \= \ave{\psi^T\,C\gamma^0\gamma_5\,\tau_2 \,\lambda_2 \,\psi}
     \;,
\label{delta2}
\end{equation}
which breaks all three symmetries, i.e., Lorentz invariance, color-$SU(3)$, 
and chiral symmetry, at the same time. It is related to a second
gap parameter $\Delta'$.

The simultaneous treatment of these six condensates leads to a set
of six coupled gap equations which we have analyzed in Ref.~\cite{BHO02a}
within an NJL-type model.
The dispersion laws for the paired quarks now take the form
\beq
    E_\mp(p) \= \sqrt{(\sqrt{\vec p^2 + M_{eff}^2}\mp\mu_{eff})^2 
    + |\Delta_{eff}|^2}~,
\eeq    
where $M_{eff}$, $\mu_{eff}$ and $\Delta_{eff}$ are functions of the
six condensates \cite{BHO02a}. It is interesting, that under certain
circumstances $\Delta_{eff}$ can vanish, even if the 
gap parameters $\Delta$ and $\Delta'$ both are non-zero. 
However, there are indications that these effectively gapless modes might 
be always unstable \cite{BHO02a}.

\begin{figure}
\centering
\includegraphics[width=6cm]{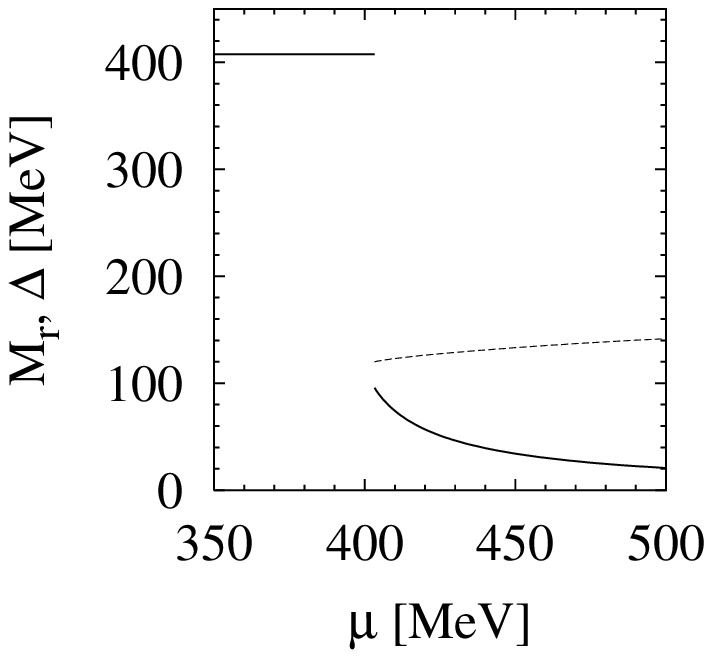}
\hfill
\includegraphics[width=6cm]{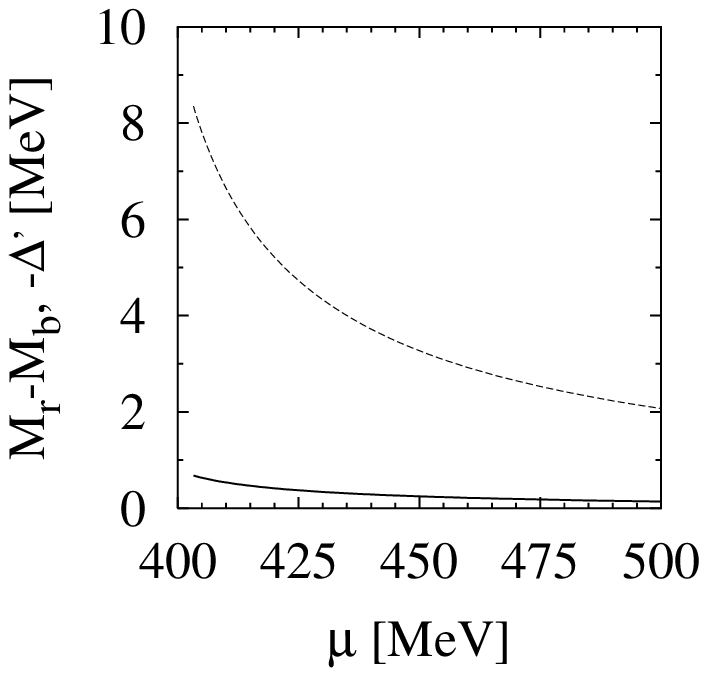}
\caption{Various quantities as functions of the quark chemical potential $\mu$
         \cite{BHO02a}.
         Left: $M_r$ (solid), $\Delta$ (dashed).
         Right: $M_r-M_b$ (solid), $-\Delta'$ (dashed).}
\label{gaps2sc}
\end{figure}

Some numerical results of our analysis at $T=0$ are displayed in 
Fig.~\ref{gaps2sc}.
On the r.h.s.  we show the constituent mass $M_r$ of the red 
quark and the diquark gap $\Delta$ as functions of $\mu$.
The bahavior is quite similar to the results of Ref.~\cite{BeRa99},
with a phase transition into a color-superconducting phase at
$\mu \simeq 400$~MeV.
On the r.h.s. the second diquark gap $\Delta'$ and the difference
between red and blue constituent quark masses, $M_r-M_b$ are shown.
We see that -- at least for our model interaction -- these quantities are
relatively small, which a posteriori justifies their usual negligence. 

\begin{figure}
\centering
\includegraphics[width=6cm]{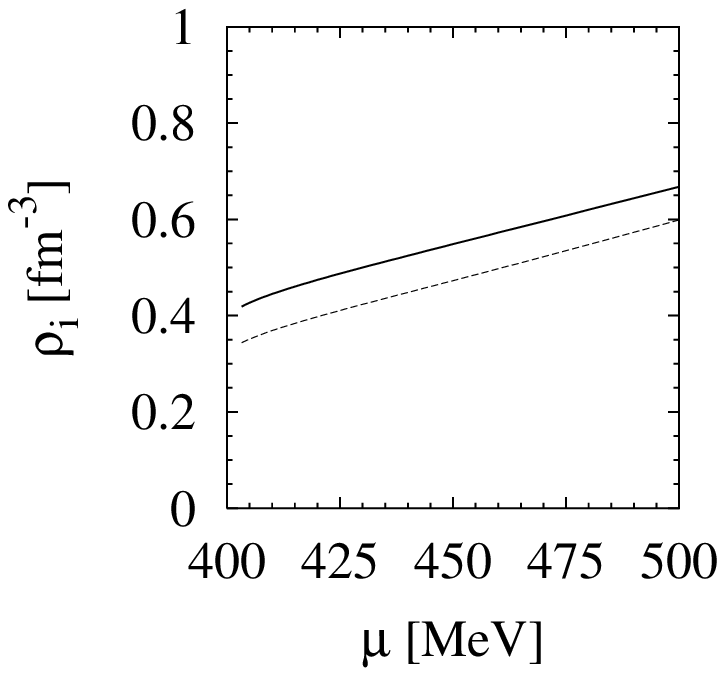}
\hfill
\includegraphics[width=6cm]{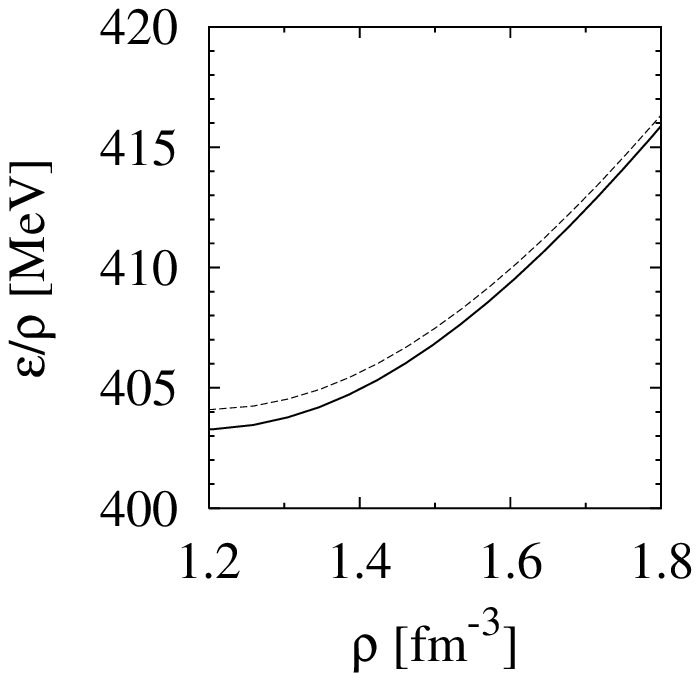}
\caption{Left: Number densities of red quarks (solid) and blue quarks
         (dashed) as functions of the quark chemical potential $\mu$.
          Right: Energy per quark as function of the total quark number
          density for a color superconducting system with equal densities
          of gapped and ungapped colors (dashed) and with unequal densities
          as given in the left panel (solid).}
\label{dens}
\end{figure}

On the l.h.s. of Fig.~\ref{dens} the number densities of red and blue
quarks are displayed as functions of $\mu$. 
As one can see, the density of the paired quarks is about
10 - 20\% larger than the density of the unpaired quarks in the regime
which is shown. One might therefore ask, how matter should arrange itself
to be color neutral.
A possible scenario could be that several domains
emerge in which the symmetry is broken into different directions,
such that the total number of red, green and blue quarks is equal.
Alternatively we could construct a uniform phase with equal densities of
all three colors. In this case the chemical potential has to be 
larger for the unpaired quarks than for the paired ones. 
In order to compare these two possibilities the energy per quark as
function of the total quark number density is shown on the r.h.s. of
Fig.~\ref{dens}. The  dashed line corresponds to quark matter with equal 
densities of paired and unpaired quarks, the
solid line corresponds to quark matter with equal chemical potentials.
Obviously the latter is energetically favored, but the difference is
small and the situation might change, if surface effects are included. 

So far we have assumed that one color (the ``blue'' quarks) does not
participate in a condensate. However, because of Cooper's theorem
we should expect that they will also condense if there is
attraction in an appropriate channel. Since only quarks of a single color 
are involved, the pairing must take place in a channel which is symmetric 
in color. Assuming $s$-wave condensation in an isospin-singlet channel, 
a possible candidate is spin-1~\cite{ARW98}.
This interesting possibility has recently been analyzed in Ref.~\cite{BHO02b}
and will be discussed in more detail in Ji\v r\'{\i} Ho\v sek's
contribution \cite{Ho02}.

\section{Three flavors}

For two flavors the flavor index in \eq{saa} was restricted to 
$A=2$. This is different for $N_f=3$, where 
the flavor operators $\tau_A$ denote the
three antisymmetric Gell-Mann matrices, i.e., $A = 2, 5, 7$,
describing $ud$-, $us$-, and $ds$-pairing, respectively.
The two-flavor condensation pattern, \eq{2sc} is still
possible, but now there are several other combinations which cannot be 
transformed into $\ct$ via color or flavor rotations.

In the case of three degenerate light flavors, dense matter 
is expected to form a so-called color-flavor locked (CFL) state
\cite{ARW99}, characterized by the situation
\beq
    \ct = \cf = \cs \neq 0  
    \quad  {\rm and}\quad
    s_{AA'} = 0 \quad {\rm if}\quad A\neq A' \,. 
\label{cfl}
\end{equation}
In this state color $SU(3)$ as well as the chiral $SU(3)_L \times SU(3)_R$
and the $U(1)$-symmetry related to baryon-number conservation are broken
down to a common $SU(3)_{color + V}$ subgroup where color and flavor 
rotations are locked.
As a consequence all gluons receive a mass and there is a gap in the
disperison laws of all nine (3 flavors, 3 colors) quark quasiparticle 
states.

The situations discussed so far are idealizations of the real world, where
the strange quark mass $M_s$ is neither infinite, such that strange quarks 
can be completely neglected, as in the previous section,
nor degenerate with the masses of the up and down quarks. 
For sufficiently large quark chemical potentials $\mu \gg M_s$, the $s$
quark mass becomes of course almost negligible against $\mu$ and matter 
is expected to be in the CFL phase.  
It is not clear, however, whether this CFL phase is directly connected to
the hadronic phase \cite{ScWi99}
at low densities, or whether an intermediate 2SC phase 
exists, where only up and down quarks are paired.  
It is obvious that the answer to this question depends on the strange quark 
mass.
This has first been analyzed by Alford, Berges and Rajagopal \cite{ABR99} 
who have studied the color-flavor unlocking phase transition in a
model calculation with different values of $M_s$. 
Assuming that the region below $\mu \simeq$~400~MeV belongs to the hadronic
phase, these authors came to the conclusion that a 2SC-phase exists if
$M_s \gsim$~250~MeV. Here $M_s$ is the constituent mass of
the strange quark, which could be considerably larger than the current
quark mass $m_s \sim$~100 to 150~MeV in the Lagrangian. Similar to the
nonstrange constituent quark masses, dicussed in the previous section,
it is in general $T$- and $\mu$-dependent and can depend on the presence of 
quark-antiquark and diquark condensates. In particular, it can be
discontinous along a first-order phase transition line.
This means, not only the phase structure depends on the effective quark 
mass, but also the quark mass depends on the phase. 

Recently, we have studied these interdependencies \cite{BuOe02,OeBu02}
within an NJL-type model defined by the Lagrangian
\beq
    {\cal L}_{eff} \= \pb (i \partial\hspace{-2.3mm}/ - \hat{m}) \psi
                      \+ {\cal L}_{q\bar q} \+ {\cal L}_{qq} \,.
\label{Lagrange}
\end{equation}
The mass matrix $\hat m$ has the form 
$\hat m = diag(m_u, m_u, m_s)$ in flavor space, where we have
assumed isospin symmetry, $m_u = m_d$.
To study the interplay between the color-superconducting
diquark condensates $s_{AA'}$ and the quark-antiquark condensates
$\phi_u$ and $\phi_s$ we consider an NJL-type interaction 
with a quark-quark part
\beq
    {\cal L}_{qq} \=
    H\sum_{A = 2,5,7} \sum_{A' = 2,5,7}
    (\pb \,i\gamma_5 \tau_A \lambda_{A'} \,C\pb^T)
    (\psi^T C \,i\gamma_5 \tau_A \lambda_{A'} \, \psi) 
    \,.
\label{Lqq}
\end{equation}
and a quark-antiquark part
\bea
   & {\cal L}_{q\bar q} \= \quad&G\, \sum_{a=0}^8 \Big[(\pb \tau_a\psi)^2
    \+ (\pb i\gamma_5 \tau_a \psi)^2\Big] 
\nonumber\\
   & \hspace{1cm}\;-\; &K\,\Big[{\rm det}_f\Big(\pb(1+\gamma_5)\psi\Big) \,+\
                   {\rm det}_f\Big(\pb(1-\gamma_5)\psi\Big)\Big]\;.
\label{Lqbarq}
\eea
As before $\tau_a, a = 1, ..., 8$, denote Gell-Mann matrices acting in 
flavor space, while $\tau_0 = \sqrt{\frac{2}{3}}\,1\hspace{-1.5mm}1_f$ is 
proportional to the unit matrix. 
\eq{Lqbarq} corresponds to a typical 3-flavor NJL-model Lagrangian.
It consists of a $U(3)_L \times U(3)_R$-symmetric 4-point interaction and a 
't~Hooft-type 6-point interaction which breaks the the $U_A(1)$ symmetry.
The latter has been neglected in Ref~\cite{BuOe02}.

Starting from this Lagrangian it is tedious, but straight forward to 
calculate the mean-field thermodynamic potential $\Omega$ at temperature 
$T$ and quark chemical potential $\mu$ (for details see Ref.~\cite{BuOe02}) and to determine the selfconsistent 
solutions for the expectation values $\phi_u=\phi_d$, $\phi_s$, $\ct$, 
and $\cf=\cs$ by minimizing $\Omega$ with respect to these expectation 
values.
In this context it is convenient to introduce the
constituent quark masses
\beq 
M_u = m_u - 4G\phi_u + 2 K \phi_u\,\phi_s~,  \qquad
M_s = m_s - 4G\phi_s + 2 K \phi_u^2~,
\label{masses}
\end{equation}
and the diquark gaps
\begin{equation}
\Delta_2 = -2H s_{22} \quad \mathrm{and} \quad \Delta_5 = -2 H
s_{55}~.
\end{equation}

To determine the values of the various condensates we first have to
specify the parameters of the interaction.  We take the parameter
values of Ref.~\cite{Rehberg} which were obtained by fitting vacuum masses 
and decay constants of pseudoscalar mesons. 
The coupling constant $H$ which cannot be fixed in this way was chosen
to yield ``typical'' values for $\Delta_2$ in the 2SC phase~\cite{OeBu02}.

\begin{figure}
\centering
\includegraphics[width=6cm]{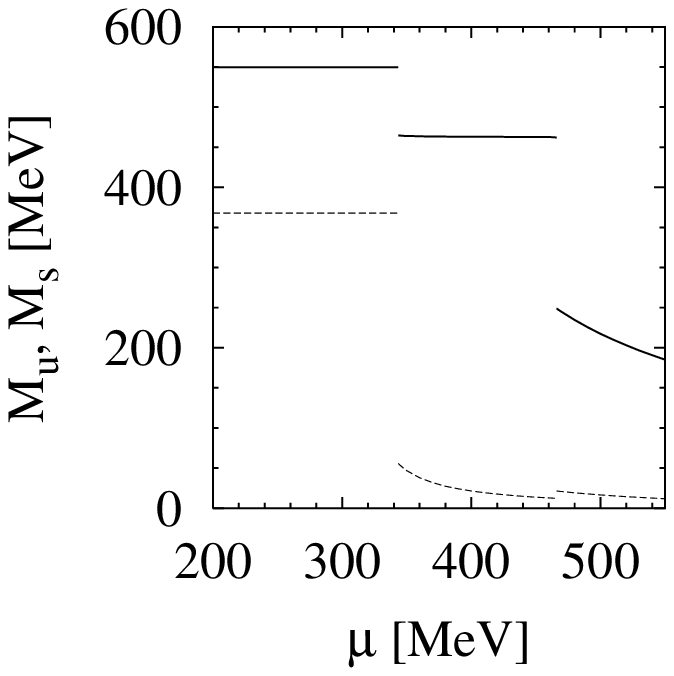}
\hfill
\includegraphics[width=6cm]{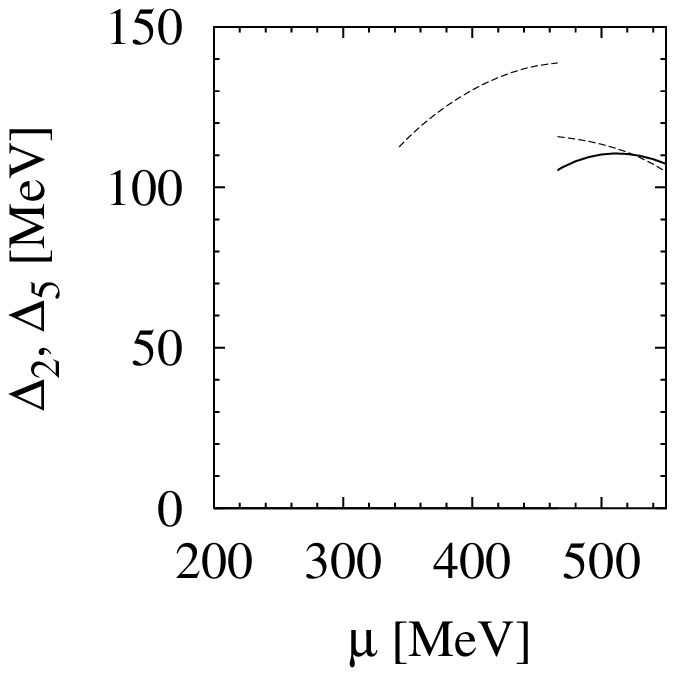}
\caption{Gap parameters at $T=0$ as functions of the quark chemical 
          potential $\mu$.
          Left: Constituent masses of up and down quarks (dashed), 
          and of strange quarks (solid).
          Right: Diquark gaps $\Delta_2$ (dashed) and $\Delta_5$ (solid).}
\label{figt=0}
\end{figure}

Our results for the constituent quark masses (left) 
and the diquark gaps (right) at $T=0$ as functions of $\mu$ 
are displayed in Fig.~\ref{figt=0}. Obviously, one can distinguish three
phases. At low $\mu$, the diquark gaps 
vanish and the constituent quark masses stay at their vacuum values.
Hence, in a very schematic sense, this phase can be identified with the 
``hadronic phase'' (although there are of course no hadrons in our model). 

At a critical $\mu = \mu_1$ a first-order phase 
transition to the 2SC phase takes place: The diquark
gap $\Delta_2$ has now a non-vanishing value, whereas $\Delta_5$
remains zero. At the same time the mass of the light quarks drops from
the vacuum value to about 50~MeV and the baryon number density jumps
from zero to about 2.5 nuclear matter density. 
At $\mu = \mu_2$ the system undergoes a second first-order phase
transition, this time from the 2SC phase into the CFL phase, which is
characterized by $\Delta_5\neq 0$ (and 
$\Delta_2 \neq 0$). 
\begin{figure}
\centering
\includegraphics[width=8cm]{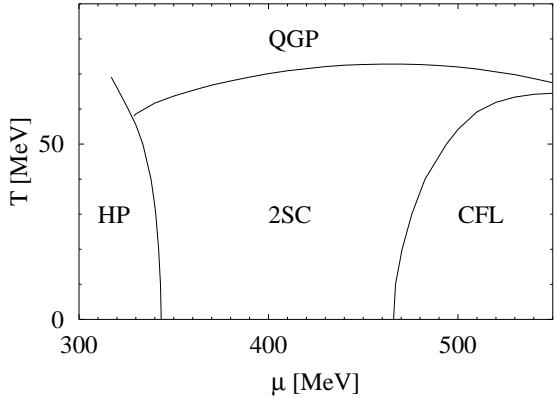}
\caption{Phase diagram in the $\mu-T$ plane.}
\label{figphase}
\end{figure}

We now extend our analysis to. $T \neq 0$. The
resulting phase diagram in the $\mu$-$T$ plane is shown in
Fig.~\ref{figphase}. 
We can distinguish four different regimes: At
low $T$ we find (with increasing $\mu$) the
``hadronic phase'', the 2SC phase, and the CFL phase.  Similar to $T = 0$
these three phases are well separated by first-order phase
transitions.  The high temperature regime is governed by the QGP
phase, which is characterized by vanishing diquark condensates and
small values of $\phi_u$ and (for large enough $\mu$) $\phi_s$.  There
we find smooth crossovers with respect to $\phi_u$ and $\phi_s$
instead of the first-order phase transitions. 
The transition from
the 2SC phase to the QGP phase is of second order and the critical 
temperature is in almost perfect agreement with the well-known BCS
relation $T_c = 0.57 \Delta_2 (T = 0)$.

It has been argued~\cite{ABR99} that the
color-flavor-unlocking transition has to be first order because
pairing between light and strange quarks can only occur if the gap is
of the same order as the mismatch between the Fermi surfaces. 
Moreover, the phase transition corresponds to a finite temperature chiral 
restoration phase transition in a three-flavor theory, and therefore the
universality arguments of Ref.~\cite{PiWi84} should apply~\cite{RaWi00}.
At low $T$  our results are in agreement with these
predictions. However, above a critical point we find a second order
unlocking transition. In fact, the above arguments are not as
stringent as they seem to be on a first sight: First the Fermi
surfaces are smeared out due to thermal effects and secondly the 2SC
phase is not a three-flavor chirally restored phase, but only
$SU(2)\times SU(2)$ symmetric.

\section{Discussion: charge neutral matter}
\label{discussion}

In this article we discussed general features of two- and three-flavor
color superconductors. For simplicity, we restricted our studies to a 
common chemical potential for all flavors. 
However, for many applications, e.g., to the decription of quark 
cores of neutron stars, one has to consider color and charge neutral
matter in $\beta$-equilibrium. 
Very recently, it was argued by Alford and Rajagopal that these constraints 
could completely rule out the existence of a 2SC phase in compact stars 
\cite{AlRa02}. This could give rise to a much larger window for 
crystalline phases than expected earlier \cite{BoRa02}.

To this end, we consider a system of massless $u$ and $d$ quarks
together with electrons, but -- in a first step -- with no strange quarks.
Since the density of electrons is small (see, e.g., \cite{BuOe99}), to 
achieve charge neutrality the density of $d$-quarks must be almost twice
as large as the density of $u$-quarks, and hence 
$\mu_d \approx 2^{1/3} \mu_u$. This means that, e.g., for $\mu_u=400$~MeV,
the Fermi momenta of $u$ and $d$ differ by about 100~MeV, making 
$ud$ BCS-pairing very difficult. Alford and Rajagopal approached the problem
from the opposite side, performing an expansion in terms of the
strange quark mass. They found that, whenever the 2SC phase is more 
favorite than no pairing at all, the CFL phase is even more favorite. 
However, this analysis did not include selfconsistently calculated
quark masses, and should be redone including these effects.
Work in this direction is in progress.


\begin{thebibliography}{99}
\bibitem{Halasz99} Halasz, M.A., Jackson, A.D., Shrock, R.E., 
                   Stephanov, M.A., and Verbaarschot, J.J.M. (1998)
                   On the Phase Diagram of QCD,
                   {\it  Phys. Rev.}, {\bf D 58}, 
                   096007
\bibitem{Raja99}   Rajagopal, K. (1999)
                   Mapping the QCD Phase Diagram,
                   {\it Nucl. Phys.}, {\bf A 661}, pp.~150--161
\bibitem{RaWi00}   Rajagopal K. and Wilczek F. (2001) 
                   The Condensed Matter Physics of QCD, 
                   in Shifman, M. (ed.), 
                   {\it B.L. Ioffe Festschrift, At the Frontier of
                   Particle Physics / Handbook of QCD}, {vol. 3},
                   World Scientific, Singapore, pp.~2061--2151
\bibitem{Alford01} Alford, M. (2001) 
                   Color superconducting quark matter,
                   {\it Ann. Rev. Nucl. Part. Sci.}, {\bf 51}, pp.~131--160
\bibitem{San02}    Sannino, F. (2002)
                   Aspects of the Quantum Chromo Dynamics Phase Diagram,
                   {\it e-Print archive}, {\bf hep-ph/0205007}, pp.~16
\bibitem{CoPe75}   Collins, J.C. and Perry, M.J. (1975) 
                   Superdense matter: Neutrons or asymptotically free quarks?,
                   {\it Phys. Rev. Lett.} {\bf 34},
                   pp.~1353--1356.
\bibitem{Ba77}     Barrois, B. (1977) 
                   Superconducting quark matter
                   {\it Nucl. Phys.}, {\bf B 129}, pp.~390--402  
\bibitem{Fr78}     Frautschi, S.C. (1978) Asymptotic freedom and color 
                   superconductivity in dense quark matter,
                   in Cabibbo, N. (ed.)
                   {\it Proc. of the Workshop on Hadronic Matter at Extreme 
                   Energy Density}, Erice
\bibitem{BaLo84}   Bailin, D. and Love, A. (1984)
                   Superfluidity and Superconductivity in Relativistic
                   Fermion Systems,
                   {\it Phys. Rep.}, {\bf 107}, pp.~325--385
\bibitem{ARW98}    Alford, M., Rajagopal, K., and Wilczek, F. (1998)
                   QCD at Finite Baryon Density: Nucleon Droplets and 
                   Color Superconductivity, 
                   {\it Phys. Lett.}, {\bf B 422} 247--256
\bibitem{RSSV98}   Rapp, R., Sch\"afer, T., Shuryak, E.V., and 
                   Velkovsky, M. (1998) 
                   Diquark Bose Condensates in High Density Matter and 
                   Instantons,
                   {\it Phys. Rev. Lett.}, {\bf  81}, pp.~53--56
\bibitem{FW71}     Fetter A.L. and Walecka, J.D. (1971)
                   Quantum theory of many-particle systems, 
                   Mc Graw-Hill, New York, pp.~601
\bibitem{PiRi99}   Pisarski, R.D. and  Rischke, D.H. (1999) 
                   Superfluidity in a Model of Massless Fermions Coupled 
                   to Scalar Bosons,
                   {\it  Phys. Rev.}, {\bf D 60}, 094013
\bibitem{We99}     Weber, F. (1999)
                   From Boson Condensation to Quark Deconfinement: 
                   The Many Faces of Neutron Star Interiors,
                   {\it Acta Phys. Polon.}, {\bf B 30}, pp~3149--3169
\bibitem{BGS01}    Blaschke, D.,  Glendenning, N.K., and Sedrakian A. (eds.)
                   (2001) Physics of Neutron Star Interiors,
                   {\it Lecture Notes in Physics}, {\bf vol. 578},
                   Springer, Berlin, Heidelberg
 \bibitem{PiRi00}  Pisarski, R.D. and  Rischke, D.H. (2000)
                   Gaps and Critical Temperature for Color Superconductivity,
                   {\it Phys. Rev.}, {\bf D 61}, 051501
                   Color superconductivity in weak coupling, ibd., 
                   074017
\bibitem{ARW99}    Alford, M., Rajagopal, K., and Wilczek, F. (1999)
                   Color-Flavor Locking and Chiral Symmetry Breaking 
                   in High Density QCD,
                   {\it Nucl. Phys.}, {\bf B 537}, pp.~443--458
\bibitem{RSZ01}    Rapp, R., Shuryak, E.V., and Zahed, I. (2001)
                   A chiral crystal in cold QCD matter at intermediate 
                   densities?, 
                   {\it Phys. Rev.}, {\bf D 63}, 034008
\bibitem{ABoR01}   Alford, M., Bowers, J., and Rajagopal, K. (2001)
                   Crystalline Color Superconductivity,
                   {\it Phys. Rev.}, {\bf D 63}, 074016
\bibitem{Sch00a}   Sch\"afer, T. (2000)
                   Kaon Condensation in High Density Quark Matter,
                   {\it Phys. Rev. Lett.}, {\bf 85},  pp.~5531-5534
\bibitem{BeSc02}   Bedaque,  P.F. and Sch\"afer, T (2002) 
                   High Density Quark Matter under Stress, 
                   {\it Nucl. Phys.}, {\bf A 697}, pp.~802-822
\bibitem{Sch00b}   Sch\"afer, T. (2000) 
                   Quark Hadron Continuity in QCD with one Flavor,
                   {\it Phys. Rev.}, {\bf D 62}, 094007
\bibitem{BHO02b}   Buballa, M., Ho\v sek, J., and Oertel, M. (2002)
                   Anisotropic admixture in color-superconducting quark 
                   matter, 
                   {\it  e-Print archive}, 
                   {\bf hep-ph/0204275}, pp.~4.
\bibitem{SanS02}   Sannino, F. and Sch\"afer, W. (2002)
                   Relativistic massive vector condensation,
                   {\it Phys. Lett. } {\bf B 527}, pp.~142-148 
\bibitem{AlRa02}   Alford, M. and Rajagopal, K. (2002)
                   Absence of two-flavor color superconductivity in compact 
                   stars ,
                   {\it  e-Print archive}, 
                   {\bf hep-ph/0204001}, pp.~17
\bibitem{Son99}    Son, D.T. (1999)
                   Superconductivity by long-range color magnetic interaction 
                   in high-density quark matter,
                   {\it Phys. Rev.}, {\bf D 59}, 094019;
\bibitem{SW99}     Sch\"afer, T. and Wilczek, F. (1999) Superconductivity from perturbative 
                   one gluon exchange in high density quark matter, 
                   {\it Phys. Rev.} {\bf D60}, 114033;
\bibitem{HMSW00}   Hong, D.K., Miransky, V.A., Shovkovy, I.A., and Wijewardhana, L.C.R. (2000) 
	           Schwinger-Dyson approach to color superconductivity in dense QCD,
                   {\it Phys. Rev.} {\bf D61}, 056001, {\it err.} {\bf D62}, 059903.
\bibitem{RaSh00}   Rajagopal, K., Shuster, E. (2000) 
                   On the Applicability of Weak-Coupling Results 
                   in High Density QCD,
                   {\it Phys. Rev.}, {\bf D 62}, 085007
\bibitem{RSSV00}   Rapp, R., Sch\"afer, T., Shuryak, E.V., and 
                   Velkovsky, M. (2000)
                   High Density QCD and Instantons,
                   {\it Annals Phys.}, {\bf 280}, pp.~35--99
\bibitem{Ri00}     Rischke, D.H. (2000)
                   Debye screening and Meissner effect in a two-flavor 
                   color superconductor,
                   {\it Phys. Rev.}, {\bf D 62}, 034007
\bibitem{BeRa99}   Berges, J. and Rajagopal, K. (1999)
                   Color Superconductivity and Chiral Symmetry Restoration 
                   at Nonzero Baryon Density and Temperature,
                   {\it Nucl.Phys.}, {\bf B 538}, pp.~215--232
\bibitem{LaRh99}   Langfeld, K. and Rho, M. (1999)
                   Quark Condensation, Induced Symmetry Breaking and 
                   Color Superconductivity at High Density,
                   {\it Nucl.Phys.}, {\bf A 660}, pp.~475-505
\bibitem{BHO02a}   Buballa, M., Ho\v sek, J., and Oertel, M. (2002)
                   Self-consistent parametrization of the two-flavor 
                   isotropic color-superconducting ground state,
                   {\it Phys. Rev.}, {\bf D 65}, 014018
\bibitem{Ho02}     Ho\v sek, J. (2002)
                   Anisotropic QCD superfluids,
                   {\it these proceedings}
\bibitem{ScWi99}   Wilczek, F. and Sch\"afer, T (1999)
                   Continuity of Quark and Hadron Matter,
                   {\it Phys. Rev. Lett.}, {\bf 82}, pp.~3956--3959
\bibitem{ABR99}    Alford, M., Berges, J., and Rajagopal, K. (1999)
                   Unlocking Color and Flavor in Superconducting Strange 
                   Quark Matter
                   {\it Nucl. Phys.}, {\bf B 558}, pp.~ 219--242
\bibitem{BuOe02}   Buballa, M. and Oertel, M. (2002)
                   Color-Flavor Unlocking and Phase Diagram with 
                   Self-Consistently Determined Strange Quark Masses,
                   {\it Nucl. Phys.}, {\bf A}, in press;
                   {\it  e-Print archive}, 
                   {\bf  hep-ph/0109095}, pp.~16
\bibitem{OeBu02}   Oertel, M. and Buballa, M. (2002)
                   Color-Flavor (Un-)locking,
                   in  Buballa, M. et al. (eds.)
                   {\it Ultrarelativistic heavy ion collisions,
                   Proc. of the International workshop XXX 
                   on gross properties of nuclei and nuclear excitations, 
                   Hirschegg}, GSI, Darmstadt
\bibitem{Rehberg}  Rehberg, P., Klevansky, S.P., and H\"ufner, J. (1996)
                   Hadronization in the SU(3) Nambu-Jona-Lasinio Model
                   {\it Phys. Rev.}, {\bf C 53}, pp.~410--429
\bibitem{PiWi84}   Pisarski, R.D. and Wilczek, F. (1984)
                   Remarks on the chiral phase transition in chromodynamics,
                   {\it Phys. Rev.}, {\bf D 29}, pp.~338--341
\bibitem{BoRa02}   Bowers, J.A. and Rajagopal, K. (2002)
                   The Crystallography of Color Superconductivity,
                   {\it  e-Print archive}, 
                   {\bf  hep-ph/0204079 }, pp.~42
\bibitem{BuOe99}   Buballa, M. and Oertel, M. (1999)
                   Strange quark matter with dynamically generated 
                   quark masses,
                   {\it Phys. Lett.}, {\bf B 457}, pp.~261--267
\end{thebibliography}
\end{document}